\documentstyle[prl,aps,epsfig,multicol,amssymb,amsfonts]{revtex}
\tighten
\renewcommand{\narrowtext} 
{\begin{multicols}{2}\global\columnwidth20.5pc} 
\renewcommand{\widetext}
{\end{multicols}\global\columnwidth42.5pc} 
\begin{document} 
\draft 
\title{Multifractality of wavefunctions at the quantum Hall transition
revisited} 
\author{F.~Evers$^1$, A.~Mildenberger$^2$,  and A.~D.~Mirlin$^{1,2,*}$ } 
\address{$^1$Institut
f\"ur Nanotechnologie, Forschungszentrum Karlsruhe, 76021 Karlsruhe,
Germany}
\address{$^2$Institut f\"ur Theorie der Kondensierten Materie,
Universit\"at Karlsruhe, 76128 Karlsruhe, Germany}
\date{\today}
\maketitle
\begin{abstract}
We investigate  numerically the statistics of wavefunction
amplitudes $\psi({\bf r})$ at the integer quantum Hall transition.
It is demonstrated that in the limit of a large system size
the distribution function of $|\psi|^2$ is log-normal, so that
the multifractal spectrum $f(\alpha)$ is exactly parabolic. 
Our findings lend strong support to a recent conjecture for a critical
theory of the quantum Hall transition. 
\end{abstract}

\narrowtext

In 1980 von Klitzing, Dorda, and Pepper discovered \cite{vonKlitzing80} that
the Hall conductance $\sigma_{\rm xy}$ of a two-dimensional electron gas 
develops plateaus at values quantized in units of $e^2/h$.
%Since then the integer quantum Hall effect (IQHE) is commonly considered as
%one of the most striking macroscopic quantum phenomena \cite{prange87}.
Although two decades have passed, the integer quantum Hall effect
(IQHE) still constitutes one of the great
challenges of condensed matter physics. 
Initially, the effort focussed on the physics of the
Hall plateaus which is by now  fairly well understood
\cite{prange87,janssen94}.
Then interest has shifted towards the transition region,
where $\sigma_{\rm xy}$ crosses over from one
plateau to the next. However, here the situation is not as well resolved.
It has been understood from early on
that this is a second-order phase transition, and the  scaling scenario
has been confirmed in numerous experiments and computer
simulations \cite{huckestein94}.
In particular, the numerical prediction for the critical index
of the localization length $\nu\simeq 2.35$ has been verified experimentally.
In contrast, analytical approaches did not lead to
quantitative predictions. The ultimate goal here is to identify
the effective low-energy theory of the critical point 
(expected to be a conformal field theory) and to calculate, on this basis,
critical exponents and other characteristics of the transition region.

%Let us give a brief account of relevant developments.
The earliest field-theoretical formulation of the problem was given
by Pruisken \cite{pruisken84} (see \cite{weidenmueller87} for a more precise,
supersymmetric version) and has the form of the non-linear
$\sigma$-model with a topological term. 
Since the latter is invisible in perturbation theory, one has to resort to
non-perturbative means in order to address the critical behavior.
Pruisken and coworkers were thus led to the dilute instanton gas
approximation \cite{levine84}. However, this approximation can only be
justified in the weak-coupling limit, $\sigma_{xx}/ (e^2/h)\gg 1$, 
and becomes uncontrolled in the critical region
$\sigma_{xx} \sim e^2/h$. For
this reason, no quantitative predictions for critical properties have been
made within this approach. 
Another line of efforts was based on a mapping of the low-energy sector of
Pruisken's model onto an antiferromagnetic superspin chain \cite{zirnbauer94}.
The superspin chain was also obtained by starting from the Chalker-Coddington
network model of the IQHE \cite{lee94}. However, attempts to find an analytical
solution of the superspin chain problem remained unsuccessful. 

Recently, two papers appeared which may signify a breakthrough in the quest
for the conformal 
critical theory of the IQHE. Zirnbauer \cite{zirnbauer99} and, a few
months later, Bhaseen {\it et al.} \cite{bhaseen00}, proposed that this is a
$\sigma$-model with the Wess-Zumino-Novikov-Witten (WZNW) term $\Gamma$,
  \begin{equation}
\label{e1}
 S[g]={1\over 8\pi\lambda^2}\int d^2x\:{\rm Str}\partial_\mu g^{-1}  
 \partial_\mu g + k\Gamma, 
%\nonumber \\
%&+&{ik\over 12\pi}\int d^3x{\rm Str}\epsilon^{\mu\nu\rho}
% g^{-1}\partial_\mu g  g^{-1}\partial_\nu g  g^{-1}\partial_\rho g,
  \end{equation}
where $g$ belongs to a certain supersymmetric superspace (see 
\cite{zirnbauer99,bhaseen00} for a detailed exposition), and $\lambda$ and $k$ are
coupling constants. The theories considered in \cite{zirnbauer99} and
\cite{bhaseen00} differ only in the value of the constant $k$ in front of the
WZNW term, $k=1$ in \cite{zirnbauer99} and $k=1/\lambda^2$ in
\cite{bhaseen00}. If true, the conjecture of
Refs.~\cite{zirnbauer99,bhaseen00} will provide a framework for a systematic
study of the IQHE critical behavior. While the exponent $\nu$ 
has not yet been found on this basis, a non-trivial
prediction for the statistics of critical eigenfunctions has been
obtained. Specifically, it was found in \cite{bhaseen00} that the
corresponding multifractality spectrum is exactly parabolic (which means that
the distribution of eigenfunction intensities is log-normal),
\begin{equation}
\label{eq:1}
\tilde f(\alpha) = 2 - (\alpha-\alpha_0)^2 / 4(\alpha_0 - 2),
\end{equation}
with $\alpha_0-2=2\lambda^2$ (a formal definition of the function
$\tilde{f}(\alpha)$ is given
below). For the model of \cite{zirnbauer99} a parabolic
multifractal spectrum of a somewhat different quantity, the two-point
conductance, was obtained \cite{janssen99}. We can show, however
\cite{note1} 
that the two statements are closely related, so that the findings of 
\cite{zirnbauer99,janssen99} imply again the result (\ref{eq:1}) for the
eigenfunction statistics, with $\alpha_0-2=4\lambda^2$. The exact parabolicity
(\ref{eq:1}) of the multifractality spectrum constitutes a stringent
prediction, 
numerical verification of which would be a serious check of validity of the
theory proposed in \cite{zirnbauer99,bhaseen00}. This is the aim of the
present paper.

Our interest in this problem was additionally motivated by the fact that the
previously published numerical results 
\cite{janssen94,huckestein94,huckestein92,janssen96,klesse95,janssen94a}
appeared to be in clear conflict with the prediction (\ref{eq:1}), showing
strong deviations from parabolicity. Furthermore, it has been a widespread
belief that the parabolic approximation cannot in principle be exact, since
the singularity spectrum $f(\alpha)$ is only defined on an interval
$\alpha_-<\alpha<\alpha_+$, where it is positive, and has infinite derivatives
at the termination points, $f'(\alpha_-)=f'(\alpha_+)=\infty$. As we
demonstrate below, these earlier findings were crucially affected by
finite-size effects and by the absence of a proper ensemble averaging.

To begin with, we recall the general framework 
used for the description of critical wavefunction statistics.
One introduces a set of inverse participation ratios \cite{note2} 
\begin{equation}
\label{e2}
P_q = \int_{L^d} d^dr |\psi({\bf r})|^{2q}, \qquad q\ge 0,
\end{equation}
where $\psi({\bf r})$ is the wavefunction amplitude and $L$ the system size.
(For the IQHE dimensionality is $d=2$.)
These moments show in the large $L$-limit 
scaling behavior characterized by a set of exponents,
which can be defined for both the average and the typical value
of $P_q$, 
\begin{eqnarray}
\label{eq:a}
&& \langle P_q \rangle = \tilde c_q L^{-\tilde \tau_q}, \\
&& P_q^{\rm typ}\sim \exp \langle \ln P_q \rangle = c_q L^{-\tau_q},
\label{eq:b}
\end{eqnarray}
where $\langle\ldots\rangle$ denotes ensemble averaging. 
As long as $q$ is small enough,
$q<q_c$, the distribution function of $P_q$ is sufficiently narrow and 
$\tilde\tau_q=\tau_q$. However, with increasing $q$ this
distribution function becomes broader. Most importantly, it shows a power-law
tail $\propto P_q^{-1-x_q}$ at large $P_q$, where $x_q$ decreases with
increasing $q$. The critical value $q_c$ is determined by the condition
$x_{q_c}=1$. For $q>q_c$ the average $\langle P_q \rangle$ is governed by
rare events, and $\tau_q > \tilde\tau_q$. We refer the reader to
Ref.\cite{mirlin00a} for more details.

Instead of using the moments $P_q$ one can study directly the distribution
$\cal P$ of the wavefunction amplitudes. It is easy to see
that (\ref{eq:a}) translates, in the limit $L\to\infty$, into
\begin{equation}
\label{eq:c}
{\cal P} ( \alpha ) \sim (\ln L)^{1/2} L^{-d +
\tilde f(\alpha)}, \ \ \  \alpha = -\ln|\psi|^2/\ln L,
\end{equation}
where the function $\tilde f(\alpha)$ is related  
to the exponents $\tilde\tau_q$ via the Legendre transformation,
\begin{equation}
\label{e3}
\tilde\tau_q = q\alpha - \tilde f(\alpha); \qquad q=\tilde
f^\prime(\alpha).
\end{equation}
Similarly, one can define the $f(\alpha)$-spectrum of a typical 
eigenfunction (which satisfies $f(\alpha)\ge 0$ \cite{janssen94}),   
\begin{equation}
\label{e4}
f(\alpha) = q\alpha - \tau_q; \qquad q=f^\prime(\alpha).
\end{equation}
The two definitions coincide, $f(\alpha)=\tilde f(\alpha)>0$ \cite{note2},  for $q<q_c$,
or, equivalently, for $\alpha_{-}<\alpha<\alpha_0$, where $\alpha_0$ 
describes the scaling of the typical value,
%\begin{equation}
%\label{eq:0}
$\exp \langle \ln |\psi|^2 \rangle \propto L^{-\alpha_0}$,
%\end{equation}
and $\alpha_{-}$ denotes the zero of $f$,
$f(\alpha_{-})=\tilde f(\alpha_{-})=0$.
%Note, however, that it can be shown \cite{janssen94} that $f(\alpha)$ is
%strictly positive, $f(\alpha)>0$, whereas $\tilde f(\alpha)$ may be
%of any sign. In particular, 
On the other hand, in the regime of rare events,
$\alpha<\alpha_{-}$, the function $\tilde f(\alpha)$ becomes negative, while
$f(\alpha)$ is not defined. The best estimate for $\alpha_0$ from
previous numerical work is $2.26\pm0.01$ \cite{klesse95}.
The value of $\alpha_{-}$ obtained in \cite{klesse95} is $1.1\pm 0.1$,
incompatible with (\ref{eq:1}).

It is important to realize that the field-theoretical prediction
(\ref{eq:1}) refers to the function $\tilde f(\alpha)$, since in the theory one
deals with averaged moments (\ref{eq:a}). In contrast, the earlier
numerical studies were devoted to the multifractal spectrum of a single
eigenfunction, thus yielding $f(\alpha)$ as an output. We further turn to a
common misconception concerning an infinite slope of the $f(\alpha)$ spectrum
at its termination point,  $f'(\alpha_-)=\infty$, which seems to rule out the
parabolic form (\ref{eq:1}). Indeed, in any {\it finite} system
$\tau_q$ must be an analytic function of $q$, implying that  $f(\alpha)$ meets
the $\alpha$-axis with infinite slope  \cite{janssen94}. This is, however,
nothing but an artefact of a finite system size $L$. For any given $L$ there
is a finite interval of $\alpha$ close to $\alpha_-$ where the corresponding
spectrum $f_L(\alpha)$ deviates from its asymptotic form and approaches the
axis with an infinite derivative. In the limit $L\to\infty$ this
interval shrinks to zero, and the spectrum acquires its limiting form
$f_\infty(\alpha)\equiv f(\alpha)$, corresponding to a non-analytic $\tau_q$,  
\begin{equation}
\tau_q = \left\{ \begin{array}{ll}
                  \tilde\tau_q,   & \qquad 0<q<q_c \\
                  q\alpha_{-},    & \qquad q>q_c.
                  \end{array}
        \right.
\end{equation}
The situation is similar to a phase transition
where  singularities also occur in the thermodynamic limit
only. Practically speaking,
this means that $\tau_q$ and $\tilde\tau_q$,
though identical in the thermodynamic limit on the interval $0<q<q_c$,
have different finite-size corrections.

%Having introduced the general formalism, we are prepared to present our
%numerical results. 
For calculating the critical wavefunctions at the
quantum Hall transition we employ the Chalker-Coddington network model
\cite{chalker87}.
%, which describes a
%quantum particle traveling along directed links of
%a square network and turning either left or right (with a probability
%1/2) whenever it reaches a
%node. The dynamics is defined by a random phase factor
%attributed to every link and by a transfer matrix characterizing the
%scattering act at a given node. 
In order to obtain the wavefunction we translate the lattice
dynamics into a unitary time evolution operator ${\mathbf U}$ 
which describes the wave 
packet propagation on the network in  discrete time steps \cite{klesse95}.
The desired critical wavefunctions are the eigenfunctions of ${\mathbf U}$,
which are found by numerical diagonalization.
%The procedure has been introduced by Metzler and Klesse,
%see Ref.~\cite{klesse95}.

The calculations were performed for systems of the size ranging from  $L=16$ 
($6\cdot 10^6$ wave functions) to  $L=1280$ ($2000$ wave functions).
We implemented efficient numerical packages \cite{num1,num2,num3} 
which allowed us not only to (partly) diagonalize large systems, but
also to do it fast enough in order to collect sufficient
statistics. Specifically, for a system of size $L$ and for each disorder
realization we diagonalized a complex $N\times N$ matrix of the size 
$N=2 L^2$ (which reached $N\approx 3.3\cdot 10^6$ for the largest $L$) 
with machine accuracy.
%One complication that had to be
%overcome is related to the fact that ${\mathbf U}$ is unitary,
%so that the eigenvalues cluster on the unit circle and are not well
%separated.
For comparison,
the biggest realization of the network model reported in
the literature that we are aware of
is smaller by a factor of five in linear dimensions \cite{klesse95}. 
Since an accurate extrapolation to the thermodynamic limit was of primary
importance for our work, it was crucial that
we could observe the finite size corrections over almost
two orders of magnitude in $L$.

As a first test of the validity of Eq.~(\ref{eq:1}), we check its two
particular implications,
namely $\alpha_0-2=2-\alpha_1$ and $\alpha_{1/2}=2$, where  $\alpha_q$ is
determined by (\ref{e3}),  $\alpha_q=d\tilde\tau_q/dq$.
We evaluated $\alpha_q(L)$ for different system sizes,
\begin{equation}
\label{eq:3}
\alpha_q(L) = -\langle |\psi|^{2q}\ln |\psi|^2\rangle
/ \langle |\psi|^{2q}\rangle \ln L,
\end{equation}
and then extrapolated to the infinite system size, 
$\alpha_q = \lim_{L\to\infty} \alpha_q(L)$,  according to 
\begin{equation}
\label{eq:3a}
\alpha_q(L) =  \alpha_q + \frac{\lambda_q}{\ln L}
\left(1+ \gamma_q {\xi_{\rm irr}^y\over L^y}+\cdots\right).
\end{equation}
The most important finite-size effect is of the form $1/\ln L$ according to
(\ref{eq:a}), with a coefficient $\lambda_q=d\ln\tilde c_q/dq$. The second
term in brackets in (\ref{eq:3a}) stems from finite-size corrections to
(\ref{eq:a}) and is governed by the leading irrelevant scaling exponent $y$. 
The corresponding length scale $\xi_{\rm irr}$
can be in principle absorbed in the coefficient $\gamma_q$.

\begin{figure}
\centerline{
\includegraphics[width=0.8\columnwidth,clip]{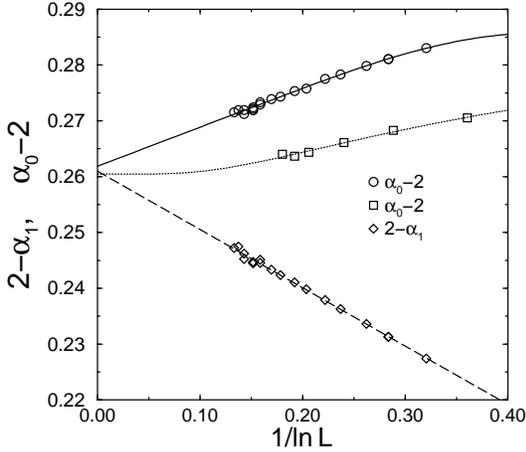}}
%\vspace{3mm}
\caption{Finite size effects for the scaling exponents $\alpha_0$
($\circ$, fit solid line) and $\alpha_1$ ($\diamond$, fit dashed line)
defined in (\ref{eq:3}). Also shown is the result for $\alpha_0$
($\Box$, fit dotted line) from the transfer matrix calculation, 
Eq.~(\ref{eq:4}).}
\label{fig:1}
\end{figure}

The raw data that provide the basis for the extrapolation to
$L\to \infty$ are shown in Fig. \ref{fig:1}.
From the fit (\ref{eq:3a})
we find $\alpha_0=2.261\pm0.003$, $\lambda_0=0.09\pm0.02$  and
$\alpha_1=1.739\pm0.002$, $\lambda_1=-0.11\pm0.01$.
A similar analysis for the case $q=1/2$ yields
$\alpha_{1/2}-2=0.0001\pm0.0002$, $\lambda_{1/2}=-0.042\pm0.002$.
It is clearly seen that within the accuracy of the numerical data
$\alpha_0-2$ and $2-\alpha_1$ indeed coincide and $\alpha_{1/2}-2$
vanishes, in agreement with (\ref{eq:1}). 
The irrelevant exponent $y$ is found to be $y=0.4\pm0.1$, which is
consistent with earlier results \cite{huckestein94,evers98} (the accuracy of
its determination is not very high, since the corresponding finite-size
correction is rather small).

As an additional check we have calculated $\alpha_0$
in yet another way, which uses a conformal mapping to
a quasi-one-dimensional strip of the Chalker-Coddington network.
It was  shown by Jan\ss en \cite{janssen94a}
that $\alpha_0$ is related to
the localization length $\xi_L$ in a strip of width $L$: 
\begin{equation}
\label{eq:4}
\alpha_0 - 2 = (1/\pi) \lim_{L\to\infty} L/\xi_L.
\end{equation}
Since this relation ties $\alpha_0$ to the amplitude $\Lambda$ of the
scaling law $\xi_L=L\Lambda$, the finite-size corrections on the strip
take a form different from (\ref{eq:3}),
\begin{equation}
\label{eq:5}
\alpha_0(L) = \alpha_0 + \gamma (\xi_{\rm irr}/L)^y + \cdots.
\end{equation}
We have calculated $\xi_L$ using the transfer
matrix method and found $\alpha_0=2.260\pm0.003$, in full
agreement with the previous result. This confirms the
fundamental assumption that the critical theory is conformally invariant.
The irrelevant scale index is obtained as
$y=0.45\pm0.1$, in agreement with earlier findings.

\begin{figure}
\centerline{
\includegraphics[width=0.75\columnwidth,clip]{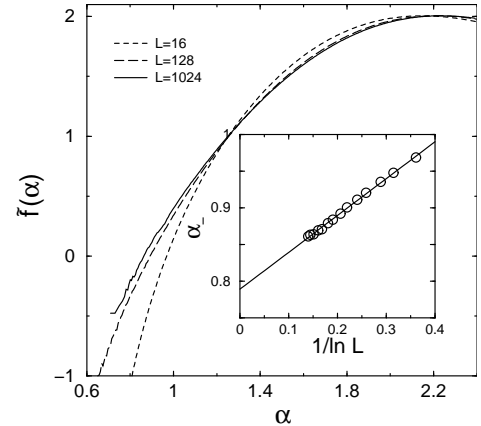}}
%\vspace{3mm}
\caption{Evolution of $\tilde f(\alpha)$, as calculated from the
distribution function  (\ref{eq:c}),
with the system size $L$. The inset shows the
corresponding change of $\alpha_{-}$, defined by
$\tilde f(\alpha_{-})=0$. An extrapolation to infinite 
$L$ yields: $\alpha_{-}=0.8\pm0.02$. }
\label{fig:3}
\end{figure}

Having satisfied ourselves that numerical results for $\alpha_0$,
$\alpha_{1/2}$, and $\alpha_1$ are in favor of the conjecture (\ref{eq:1}), we
turn to an analysis of the $\tilde f(\alpha)$ spectrum in a broader
range of $\alpha$. Results of the calculation of  $\tilde f(\alpha)$  
according to Eq.~(\ref{eq:c}) from the distribution function 
${\cal P}(|\psi^2|)$ for different system sizes are shown in
Fig.~\ref{fig:3}. It is seen that $\tilde f(\alpha)$ changes appreciably with
$L$, which points to the importance of the extrapolation to
$L\to\infty$. In particular, the inset of Fig.~\ref{fig:3} shows
the evolution of the zero $\alpha_-$  of $\tilde f(\alpha)$ with $L$.
Extrapolating to infinite system size (by taking into 
account the leading finite-size correction, which has the same $1/\ln L$ form
as in Eq.~(\ref{eq:3a})) yields the value $\alpha_{-}=0.8\pm0.02$ 
in the thermodynamic limit. This agrees with the result 
$0.816\pm0.004$ based on Eq.~(\ref{eq:1}) in combination with the value of
$\alpha_0$ found above,  providing a further strong support to the
parabolic law (\ref{eq:1}).

\begin{figure}
\centerline{
\includegraphics[width=0.8\columnwidth,clip]{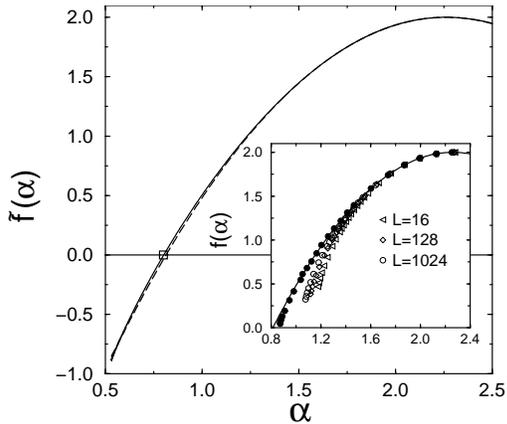}}
%\vspace{3mm}
\caption{Multifractal spectrum  $\tilde f(\alpha)$
(solid line). The dashed line shows the theoretical conjecture
(\ref{eq:1}) with $\alpha_0=2.262$. The point where $\tilde f(\alpha)$
vanishes according to the data presented in Fig. \ref{fig:3} is also
shown ($\Box$). Inset: spectrum  $f(\alpha)$ of a typical
eigenfunction for different $L$ and extrapolated to $L\to\infty$
($\bullet$). The line shows the parabolic law.}
\label{fig:4}
\end{figure}

In order to obtain the whole $\tilde f(\alpha)$ curve in the thermodynamic
limit with highest possible accuracy, we return to the procedure based on
the evaluation of moments, Eqs.~(\ref{eq:a}), (\ref{eq:3}), since it has the
advantage that we have full control over finite-size corrections, 
Eq.~(\ref{eq:3a}) (the transformation from the moments to the distribution function
Eq.~(\ref{eq:c}) induces additional corrections proportional to higher powers
of $1/\ln L$). Performing an extrapolation to $L\to\infty$, we get the 
$\tilde f(\alpha)$ curve shown in Fig.~\ref{fig:4}, which represents the
central result of this Letter. The figure demonstrates a perfect agreement
between the obtained $\tilde f(\alpha)$ and the parabolic form  (\ref{eq:1})
with $\alpha_0=2.262$. Of course, a numerical analysis can never guarantee
that two quantities are identical rather than just very close to each
other. However, taking into account that there is no small parameter in the
problem, we believe that an accidental closeness of $\tilde f(\alpha)$ to a
parabola with such a high accuracy is extremely improbable. We thus conclude
that the theoretical prediction  (\ref{eq:1}) stating that the multifractal
spectrum is exactly parabolic is correct. 

To illustrate the difference and the relation between $f(\alpha)$ and 
$\tilde f(\alpha)$ and to make a closer contact to earlier 
works, we show in the inset of Fig.~\ref{fig:4} the $f(\alpha)$
spectrum as obtained for different system sizes. The extrapolation to
$L\to\infty$ gives a good agreement with the parabola at
$\alpha>\alpha_0$, except for a close vicinity of $\alpha_-$, where
the finite-size corrections to $f(\alpha)$ become especially strong.

Finally, we mention an open issue concerning the values of the coupling
constant $\lambda$ and $k$ in Eq.~(\ref{e1}) determining $\alpha_0$ in the
theory. While Zirnbauer \cite{zirnbauer99} gave convincing arguments in favor of
$k=1$, the mechanism of fixation of $\lambda$ remains unclear. 

In conclusion, we have studied the statistics of critical wave
functions at the IQHE transition point. 
%using the Chalker-Coddington network model.
We implemented a powerful algorithm 
which allowed us to reach unprecedented large system sizes and to gather
sufficiently good statistics. Having performed the ensemble averaging and an
analysis of the finite-size corrections, we calculated the multifractality
spectrum  $\tilde f(\alpha)$ of critical wavefunctions in the thermodynamic
limit $L\to\infty$ for $0.5\le \alpha\le 2.5$. 
The result is in perfect agreement
with the parabolic form (\ref{eq:1}) with $\alpha_0=2.262 \pm 0.003$ 
and hence it supports the recent
conjecture for the form of the conformal field theory of the IQHE critical
point ~\cite{zirnbauer99,bhaseen00}.

We are thankful to X.S.~Li and to the Computer Centers of the
Forschungszentrum and the University of Karlsruhe for technical
assistance. Discussions with
M.~Jan{\ss}en,  R.~Klesse, D.G.~Polyakov,  L.~Schweitzer,
A.M.~Tsvelik, P. W\"olfle  and M.R.~Zirnbauer
are gratefully acknowledged. This work was
supported by the SFB 195 and the Schwerpunktprogramm 
"Quanten-Hall-Systeme"  der Deutschen Forschungsgemeinschaft.

\end{multicols}

\vspace{-5mm}

\begin{references}
\vspace{-20mm}
\bibitem[*]{byline}Also at Petersburg Nuclear Physics Institute,
188350 St.~Petersburg, Russia.
\bibitem{vonKlitzing80} K.~von~Klitzing {\it et al.}, 
%G.~Dorda, and M.~Pepper,
                        Phys. Rev. Lett. {\bf 45}, 494 (1980).
\bibitem{prange87} R.E.~Prange, S.~M. Girvin (eds.), {\it The Quantum Hall
                   Effect} (Springer, New York, 1992).
\bibitem{janssen94} M. Jan\ss en, O. Viehweger, U. Fastenrath, and J. Hajdu,
                    {\it Introduction to the Theory of the Integer Quantum
                      Hall Effect} (VCH, Weinheim, 1994).
%\bibitem{khmel83}      D.E.~Khmel'nitskii, JETP. Lett. {\bf 38}, 553 (1983).
\bibitem{huckestein94} B.~Huckestein, Rev. Mod. Phys. {\bf 67}, 357 (1995).
\bibitem{pruisken84}   A.M.M.~Pruisken, Nucl. Phys. B {\bf 235}, 277 (1984).
\bibitem{weidenmueller87}  H.A.~Weidenm\"uller, Nucl. Phys. B {\bf 290}, 87
                           (1987). 
\bibitem{levine84}     H.~Levine {\it et al.}, 
%S.B.~Libby, and A.M.M.~Pruisken,
                       Nucl. Phys. B {\bf 240}, 30, 49, 71 (1984).
\bibitem{zirnbauer94} M.R.~Zirnbauer, Ann. Physik {\bf 3}, 513 (1994).
\bibitem{lee94}        D.H.~Lee, Phys. Rev. B {\bf 50}, 10788 (1994);
                       J.~Kondev and J.B.~Marston, Nucl. Phys. B {\bf 497}, 639
                       (1997);
                       M.R.~Zirnbauer, J. Math. Phys. {\bf 38}, 2007 (1997). 
\bibitem{zirnbauer99}  M.~Zirnbauer, hep-th/9905054v2. 
\bibitem{bhaseen00}    M.J.~Bhaseen {\it et al.}, Nucl. Phys. B {\bf
                       580}, 688 (2000).
\bibitem{janssen99}    M.~Jan\ss en {\it et al.}, 
%M.~Metzler, M.R.~Zirnbauer,
                       Phys. Rev. B {\bf 59}, 15836 (1999). 
\bibitem{note1} The connection is established as follows. Using the
results of R.~Klesse and M.~Zirnbauer, cond-mat/0010005, it is easy to
show that $\langle T_{lm}^q \rangle \sim L^d\langle |\psi_m|^{2(1-q)}
|\psi_l|^{2q} \rangle$, where $l$ and  $m$ are two points of the network,
$T_{lm}$ is the corresponding conductance, $\psi_l$ and $\psi_m$ the
wavefunction amplitudes, and $q<1/2$. This leads to the following relation
between the exponents $X_q$ characterizing the scaling of moments of the two-point
conductance, $\langle T_{lm}^q \rangle \sim L^{-X_q}$, and the indices
$\tilde{\tau}_q$ defined in (\ref{e2}), 
$X_q=d+\tilde{\tau}_q+\tilde{\tau}_{1-q}$. The parabolic form (\ref{eq:1}) of
the wavefunction multifractality leads thus to a parabolic spectrum for the
conductance too, $X_q=X_tq(1-q)$ (which is exactly what was found in 
\cite{janssen99,zirnbauer99}), with $X_t=2(\alpha_0-d)$. 
The result $X_t=8\lambda^2$ of \cite{zirnbauer99} translates in this context
into $\alpha_0-2=4\lambda^2$.

\bibitem{huckestein92} B. Huckestein, B. Kramer, L. Schweitzer,
                       Surf. Sci. {\bf 263}, 125 (1992).
\bibitem{janssen96}    I.~Varga {\it et al.}, Europhys. Lett. {\bf 36}, 437
                       (1996);
\bibitem{klesse95}     R.~Klesse and  M.~Metzler, Europhys. Lett. {\bf 32},
                       229 (1995); Int. J. Mod. Phys. C {\bf 10}, 577 (1999)
%\bibitem{gammel94}     B.~M. Gammel, Thesis, TU Munich, 1994.
\bibitem{janssen94a}   M.~Jan\ss en, Int. J. Mod. Phys. B {\bf 8}, 943 (1994).
\bibitem{note2} We restrict ourselves to $q\ge 0$ (corresponding to
$\alpha<\alpha_0$) in this paper. One can generalize the whole consideration
to the range of negative $q$ ($\alpha>\alpha_0$), but this requires some
modification of the definition (\ref{e2}).

\bibitem{mirlin00a}    F.~Evers and A.D.~Mirlin, Phys. Rev. Lett. {\bf 84},
                       3690 (2000); 
                       A.D.~Mirlin and F.~Evers, Phys. Rev. B {\bf 62},
                       7920 (2000).
\bibitem{chalker87}    J.T.~Chalker and P.D.~Coddington, J. Phys. C
                       {\bf 21}, 2665 (1988).
%\bibitem{klesse99}     R.~Klesse and  M.~Metzler, Int. J. Mod. Phys. C {\bf
%    10}, 577 (1999). 
\bibitem{num1} P.R. Amestoy {\it et al.}
               Comput. Methods in Appl. Mech. Eng. {\bf 184}, 501 (2000);
		P.R. Amestoy  {\it et al.}, SIAM J. Matrix Anal. Appl. 
               {\bf 23}, 15 (2001).
\bibitem{num2} J. W. Demmel {\it et al.}, SIAM J. Matrix Anal. Appl.
               {\bf 20}, 720 (1999).
\bibitem{num3} R.B. Lehouca, D. Sorensen and C. Yang, {\it ARPACK
               Users Guide} (SIAM, Philadelphia, 1998).
\bibitem{evers98} F.~Evers and W.~Brenig, Phys. Rev. B {\bf 57}, 1805
  (1998). 
\end{references}
\end{document}